\documentclass[prl,twocolumn]{revtex4}
\usepackage{epsfig}
\usepackage{float}

\newcommand{\be}{\begin{equation}}
\newcommand{\ee}{\end{equation}}
\newcommand{\bea}{\begin{eqnarray}}
\newcommand{\eea}{\end{eqnarray}}

\newcommand{\p}{\partial}

\newcommand{\rd}{\mbox{d}}

\newcommand{\re}{\mbox{e}}

\newcommand{\pll}{\parallel}

\begin{document}

% \twocolumn[\hsize\textwidth\columnwidth\hsize\csname
% @twocolumnfalse\endcsname

\title{Thermodynamics of  the Double-Layer Quantum Hall Systems}

\author{Emiliano Papa$^1$ and  Alexei M. Tsvelik$^2$ }
\address{
$^1$Department of Physics, The University of Texas, Austin, TX 78712\\
$^2$Department of Physics, Brookhaven National Laboratory, Upton, NY 11973-5000, USA}
\date{\today}

\begin{abstract}

In this paper we apply the exact solution of the sine-Gordon model to describe thermodynamic properties of the  
soliton liquid  in the incommensurate phase of the double-layer quantum Hall systems.  
In this way we include thermal fluctuations and extend to finite temperatures the  results obtained  
 by C.B. Hanna, A.H. MacDonald 
and S.M. Girvin \lbrack Phys. Rev. B {\bf 63}, 125305 (2001)\rbrack. 
In addition we calculate the specific heat of the system.
 While the results obtained for the sine-Gordon model are available in a temperature interval 
$(0,T_c)$, where $T_c=8\pi \rho_s$, $\rho_s$ the pseudospin stiffness, 
they can be applied in the bilayer system up to temperatures 
$3T_{\rm BKT}$, where $T_{\rm BKT}=\pi\rho_s/2$ is the vortex mediated Berezinskii-Kosterlitz-Thouless
transition
temperature. Above this temperature the operators $\cos\beta \varphi$ and 
$\cos(2\pi \vartheta/\beta)$ are both relevant and the system is in a phase with coexisting order 
parameters. $\vartheta$ is the dual field of $\varphi$ and $\beta$ is the sine-Gordon coupling constant.
We provide numerical estimates for thermodynamic quantities for the range of parameters relevant 
for  GaAs. 

{\rm PACS No: 71.10.Pm, 65.40.-f, 65.50.+m }
\end{abstract}
% \vskip2pc]

\maketitle

% \vspace{6mm}

 Strong electron-electron interactions in  Quantum Hall Double Layer systems lead  to interesting collective effects\cite{yang}. 
 Perhaps, the most remarkable among them  is the interlayer coherence which  is established when the interlayer spacing $d$ is of the order of the magnetic length. In that  case 
the Coulomb interaction between electrons on different layers is comparable
to the interaction on  the same layer  \cite{girvin}. 
 The dynamics of this coherent state is conveniently described in terms of the  pseudospin variable for which the up and down states refer to the electron
being on  the upper or the lower layer \cite{macd}, whereas real spins are  totally frozen.

 At finite interlayer distance the symmetry in  pseudospin space is lowered down to U(1) such that the corresponding Ginzburg-Landau free energy is of 
 an easy-plane anisotropic ferromagnet. 
The latter symmetry is further broken by interlayer tunneling. In the limit of strong anisotropy one can 
treat pseudospins as belonging to a plane such that at low temperatures the  free energy is simplified
down to the classical sine-Gordon model  \cite{yang}:
\bea
\label{sg-class}
\frac{F}{T} & = & \frac{1}{ T}\int \rd x \rd y
\left[ \frac{1}{2} \rho_s (\nabla {\phi} - {\bf Q})^2 
\right.
\\[2mm] \nonumber
&&\left. \hspace{20mm}+ \frac{t}{2\pi l^2}(1 - \cos{\phi})\right] \quad ,
\eea
where $(\phi -\bf{Q}\bf{r})$ is the phase angle describing pseudospins in the $XY$ plane at 
different positions. ${\bf Q}= (2\pi d /\phi_0) \, {\bf B}_{\pll} \times {\bf \hat{z}}$ 
is the parallel magnetic field wave vector, with $\phi_0=hc/e$ being one flux quantum 
between the layers. $t$ is the tunneling energy 
which generally depends on magnetic field $Q$ as well as on $m_z$
\bea
t = t_0 e^{-Q^2 l^2 /4} \sqrt{1-m_z^2} \quad .
\eea
$m_z=\nu_1-\nu_2$ is the layer imbalance and $\rho_s$ is the pseudospin
stiffness which depends on $m_z$ as well
\bea
\rho_s=(1-m_z^2)\rho_E \quad .
\eea
$\rho_s$ arises from the loss of the Coulomb exchange in the presence of a phase gradient (in what follows  we neglect the dependence of $\rho_s$ on $m_z$, considering that the 
layer imbalance is small or equivalently $\nu_1=1-\nu_2\approx1/2$).
In the above formulas $l$ is the magnetic length equal to
$(\hbar c/eB_\perp)^{1/2}$, where $B_\perp$ is the strength of the magnetic field
perpendicular to the layers.

At higher temperatures Eq.~(\ref{sg-class}) does not properly describe the pseudospin system as it does not take into account the $2\pi$ periodicity of the field $\phi$.
This free energy should be supplemented by the contribution originating from the  
vortex configurations of angle variable $\phi$: 
\bea
F_{\rm vortex}/T  = \re^{-S_0}\int \rd^2x \cos  2\pi \theta, ~~ 
{\frac{T}{\rho_s}}\p_{\mu}\theta = \epsilon_{\mu\nu}\phi \; , 
\label{dual}
\eea
where $S_0$ is the thermodynamic action of a vortex and $\theta$ is the dual field of $\phi$. 
The scaling dimensions of this term and the tunneling term are respectively 
\bea
d_{\rm v} = \frac{\pi \rho_s}{T}\quad , \quad d_{\rm t} = \frac{T}{4\pi\rho_s} \label{ds} \quad ,
\eea
such that 
\bea
d_{\rm v}d_{\rm t} = 1/4 \quad .
\eea

 In what follows we shall apply non-perturbative results derived for the sine-Gordon model to the bilayer systems. 
Strictly speaking, this approach is valid  only below the Berezinskii-Kosterlitz-Thouless transition 
temperature
\bea
T_{\rm BKT} = \frac{\pi \rho_s}{2}
\label{BKT}
\eea
when $d_{\rm v} > 2$. We shall first discuss the region $T < T_{\rm BKT}$ and then briefly discuss what happens
at higher temperatures.

To study two-dimensional  {\it classical}\, sine-Gordon (SG) model (\ref{sg-class})  we use the well known 
analogy with the {\it quantum} (1+1)-dimensional  SG model, at $T=0$.
The action of the {\it quantum} SG model  corresponding 
to (\ref{sg-class}) is:
\bea
S=\int \rd \tau \rd x\left[\frac{1}{2} (\nabla \varphi - {\bf h} \beta
/2\pi)^2 + 2 \mu (1-\cos\beta \varphi)\right] \, ,
\label{sg-quant}
\eea
where 
$\beta^2= T/\rho_s$, $\varphi={\phi}/\beta$, $\mu=t/(4 \pi l^2  T)$.
The field $|{\bf h}|=2 \pi |{\bf Q}|/\beta^2$, coupled to the soliton
topological charge density, plays the role of a  chemical potential. 
The two models have the same partition function. The current field theory description is valid only 
at distances much greater than the magnetic length $l$ which serves as the ultraviolet cutoff. 

 The ground state energy ${\cal E}_0$ of the quantum model is related to the free energy of the classical model:
\bea
{\cal E}_0 = \frac{F}{T} \quad . 
\eea
This is a fundamental formula which we use to establish a link between the exact solution of the quantum sine-Gordon model and the model describing Quantum Hall double layer.

The SG model, given by Eq.~(\ref{sg-class}), contains two competing periodicities: the periodic
potential
tends to lock the system into a commensurate configuration with the field
$\phi$ being locked  to a  minima of the cosine 
potential,  whereas the gradient  term prefers the field configuration be 
 ${\phi}={\bf Q x}$. This competition takes place only for temperatures smaller than the critical  
temperature 
\bea
T_{c}= 8\pi\rho_s  \quad , 
\eea
below which  the cosine potential is relevant. Notice that $T_c$ is 16 times larger than 
$T_{\rm BKT}$ (\ref{BKT}) and therefore the interlayer tunneling is highly relevant in the entire area
of validity of the sine-Gordon description.  In this area there is a critical value of chemical potential $Q_c(T)$ such that for $Q>Q_c$,  the competition is  resolved by a formation of
a liquid of  solitons (domain walls). Each soliton interpolates between minima of the cosine potential. 
The transition into such incommensurate state 
occurs when the soliton energy 
equals the chemical potential.

 According to  \cite{hanna}, numerical values of the parameters for a typical GaAs double layer QH sample are the following.  
The  effective mass is $m^* \approx 0.07m_e$, total particle aerial density is 
$n_T=1.0\times 10^{11} \; {\rm cm}^{-2}$, layer (midwell to midwell) 
separation is 
$d=20  \; {\rm nm}$ and tunneling energy is $t_0= 0.1$ meV. For such a sample 
we would have $l\approx 12.6$ nm, $d/l\approx 1.6$, $\hbar \omega_c \approx 6.9$ meV for 
$\nu_T=1$, and $e^2/4\pi\epsilon l \approx 8.8$ meV. This  gives 
$\rho_E \approx 0.03$ meV, which corresponds to $T_{\rm BKT} \approx 0.5K$.

At $Q=Q_c$  a single 
soliton is introduced in the system. The value of $Q_c$ as determined by the exact solution \cite{zam} is given by the equation 
\bea
lQ_c\hspace{-1mm}=\hspace{-1mm} 4 \tau h_c \hspace{-1mm}=\hspace{-1mm}
 \frac{8 \tau}{\sqrt{\pi}}
\frac{  \Gamma\left(\frac{1}{2} \frac{\tau}{1-\tau}\right)   }
     {  \Gamma\left(\frac{1}{2}\frac{1}{1-\tau}\right)   }
\left[ \mu_c
\frac{  \Gamma\left(1-\tau \right)  }
     {  \Gamma\left(1+\tau \right)          }
\right]^{1/[2(1-\tau)]} \hspace{-5mm} , \label{Q}
\eea
where
\bea
\tau= \frac{\beta^2}{8\pi}= \frac{T}{T_c} \quad , \quad 
\mu_c= \frac{t_0 e^{-Q_c^2 \, l^2/4} }{32 \pi \rho_s} \quad , \label{mu}
\eea
In the limit $T \rightarrow 0$ this equation coincides with the expression used by  Hanna {\it et al.} \cite{hanna}:
\bea
Q_cl= \frac{4}{\pi} \sqrt{\frac{t}{2 \pi  \rho_s}} \quad \quad , \quad (T=0)
\quad . \label{t0}
\eea

 $Q_c(T)$ is a monotonously decreasing function of $T$, as shown on Fig. 1. The maximal value of $Q_c$ is achieved for $T =0$; substituting the numerical values used for GaAs, we obtain the estimate for a critical magnetic field necessary to observe the soliton liquid:
\bea
Q_c(T=0)l \approx 0.3 \quad, \quad B_c(T=0) \sim 0.1 \phi_0/ld \quad . 
\eea
 From this estimate we also see that since the value of $(Q_cl/2) < 0.15$
is always small, it is not always necessary to take into account the
field dependence of the tunneling integral in Eq.~(\ref{Q}). 
(This term is however important in cases of strong magnetic fields and mistakenly it is not accounted in 
\cite{hanna} in all asymptotic cases $Q\rightarrow +\infty$ they considered. The correction terms to 
quantities like ${\bf M_{||}}$, $\chi_{\rm sol}$ etc., resulting from the $Q$ dependence of $t$, should decrease exponentially when $Q\rightarrow +\infty$ instead of the power low decay of \cite{hanna}).

 The authors of \cite{hanna} used $T \rightarrow 0$ limit to describe the soliton state. To establish whether such description can be extended to  finite $T$, we have to recall some fundamental facts about the sine-Gordon model.

 The particle spectrum 
of the quantum SG model consists of solitons  and antisolitons
 of mass $M_s$ and for $\tau < 1/2$ also of their bound 
states  
\be
M_n = 2M_s \sin\left( n \frac{\tau}{2(1 - \tau)}\right) \; ,
\; n = 1,\cdots, \left[\frac{1}{\tau}-1\right] \; ,\label{masses}
\ee
where 
\bea
M_sl \hspace{-1mm}=\hspace{-1mm}\frac{2}{\sqrt{\pi}}
\frac{  \Gamma\left(\frac{1}{2} \frac{\tau}{1-\tau}\right)   }
     {  \Gamma\left(\frac{1}{2}\frac{1}{1-\tau}\right)   }
\left[ \mu(Q)
\frac{  \Gamma\left(1-\tau \right)  }
     {  \Gamma\left(1+\tau \right)          }
\right]^{1/[2(1-\tau)]} \hspace{-5mm} ,
\eea
where 
\bea
\mu(Q) =  \frac{t_0 e^{-Q^2 \, l^2/4} }{32 \pi \rho_s} \quad .
\eea
Recall that the sine-Gordon description is valid below $T_{\rm BKT}$ which corresponds to $\tau < 1/16$.
\begin{figure}
\unitlength=1mm
\begin{picture}(80,55)
\put(-1.5,3){\epsfig{file=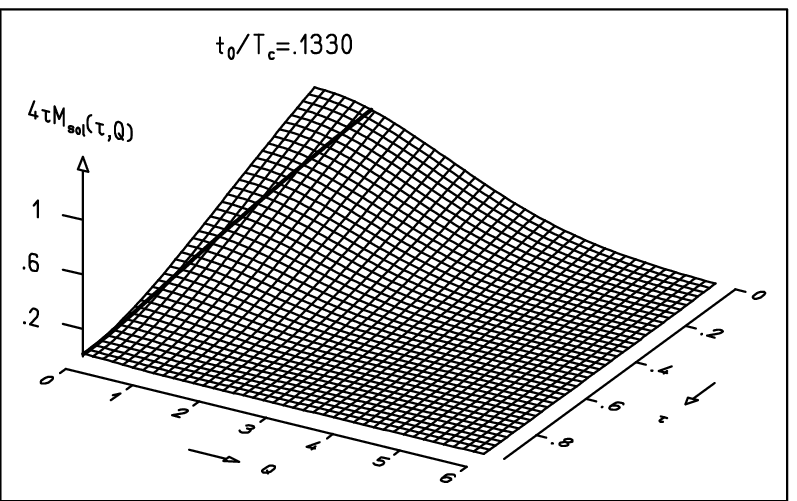,height=50mm}}
\put(50,33.8){\epsfig{file=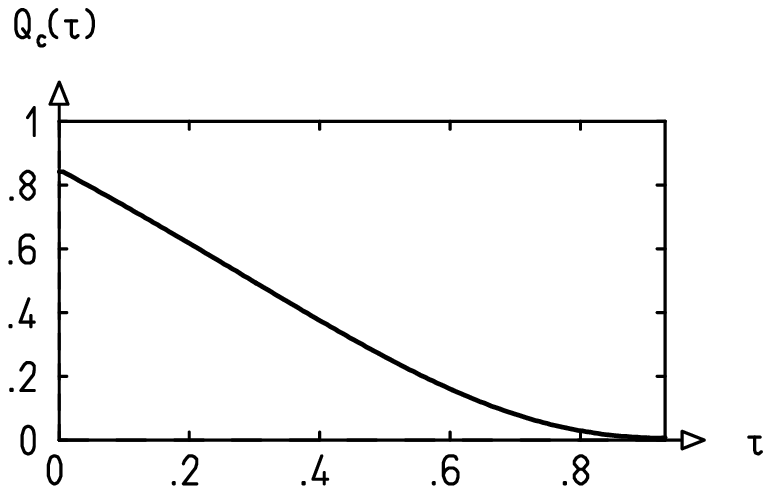,height=17.5mm}}
\put(58.3,38.0){\makebox(0,0)[cc]{{\small{\small{\small{\small C }}}}}}
\put(65.3,43.0){\makebox(0,0)[cc]{{\small{\small{\small{\small I }}}}}}
\put(54.,36){\dashbox{0.2}(0,11.1)}
\end{picture}
\caption{Plot of $4\tau M_{\rm sol}$ on temperature ($\tau=T/T_c$) and magnetic field $Q$ dependence
is shown.  $t_0$ is the tunneling energy and $T_c$ is the critical temperature above which the soliton generating operator becomes irrelevant. The line on the surface shows the critical magnetic field, above
which the system crosses to the incommensurate phase. (On the vertical axis we
show $\tau M_{\rm sol}$ instead of the diverging soliton mass $M_{\rm sol}$). As can be seen,
for fixed $\tau$, the mass of the solitons decreases exponentially as $Q$ increases. The
$Q$ and
$M_s$ axes have units $1/l$, (the inverse magnetic length, $l=12.6$ nm).}
\label{sol_mass}
\end{figure}
In the case the system will be exposed to an external "magnetic field" only the
solitons'  and antisolitons'
energy will be affected. The breathers have zero charge and do not interact
with external magnetic fields. The solitons acquire additional energy $-h$
whereas the antisolitons $h$, and in the ground state only solitons can appear.
The soliton's excitation spectrum 
(in the presence of the field $h$ coupled to the soliton charge),
is given by the following equations
\bea
\label{int-equ}
E(\theta) & = & \epsilon(\theta)=\left[\epsilon_0(\theta) - h \right]-\int_{-B}^{+B}
\rd \theta' G(\theta-\theta') \epsilon(\theta') 
\\ [3mm] \nonumber
P(\theta) &=& 2 \pi \int_0^{\theta} \hspace{-2mm} \rd \theta'\sigma(\theta')
= P_0(\theta) - \int_{-B}^{+B}  \hspace{-2mm} \rd
\theta' \vartheta(\theta-\theta') \sigma(\theta') \, . 
\eea
$\epsilon(\theta)$ is the dressed energy whereas $\epsilon_0(\theta)= M_s \cosh \theta$
and $P_0(\theta)=M_s \sinh\theta$ are the bare energy and momentum.
$\sigma(\theta)$ is the rapidity distribution function (in rapidity space).
In the above equation the boundary $B$ is defined
by the condition $\epsilon(\pm B)=0$, whereas the kernel $G(\theta-\theta')$
is the derivative with $\theta$ of the soliton-soliton scattering matrix
$S_{ss}(\theta-\theta')$.
The Fourier transform of $G(\theta-\theta')$ 
is given in Ref. \cite{zam}.

$\epsilon(\theta)$ will have a negative part only if $h\ge h_c= M_s$, where
$M_s$ is the soliton mass. This constitutes also the condition set on $h$ in order the
system to acquire a single soliton.
Therefore the relationship between the soliton mass and the critical field is:
\bea
Q_c = 4\tau M_s(Q_c) \quad . 
\eea
In the parameter range relevant for GaAs we have $\mu(0) \approx 0.03$.

In Fig.~\ref{sol_mass} we plot $\tau M_s$ as function of $\tau$ and $Q$. The soliton mass
$M_s$ diverges as $1/\tau$ as $\tau \rightarrow 0$.
On the surface of the three dimensional plot $(\tau,Q,4 \tau M_s)$ we show also the critical 
field $Q_c(\tau)$. Its dependence on temperature is shown also in the inset of Fig.~\ref{sol_mass}.
The vertical dotted line in it separates the temperature intervals $0 < T < T_{\rm BKT}$ and
$ T_{\rm BKT} < T  < T_c$.

 At $Q \neq 0$ only the solitons'  and antisolitons' 
energies are affected. The breathers have zero topological charge and do not interact
with external magnetic fields.

As we have said, the sine-Gordon model description is rigorously valid only below $T_{\rm BKT}$; however one can use it as a good approximation even above $T_{\rm BKT}$  provided the correlation length generated by the interlayer tunneling is shorter than the correlation length generated 
by vortices:
\bea
M_1(T)  \gg  [\exp(- S_0)]^{1/(2 - d_{\rm v})} \quad ,
\eea
where $M_1$ is the mass of the first breather determined by Eq.(\ref{masses}). 
Assuming that $S_0 \approx 2\pi\rho_s/T$ which is the  thermodynamic action of a skyrmion in the O(3) nonlinear sigma model, we estimate that this inequality is reasonably fulfilled up to $T \approx 3T_{\rm BKT}$.

 There are two experimentally measurable quantities one can extract from the 
sine-Gordon thermodynamics: the specific heat and the magnetization.
Both quantities are related to the free energy; at $\tau < 1/2$ the latter one   
 contains two contributions:
\bea
&&F = F_1 + F_2 \quad , \\ [3mm]
&&F_1 = \frac{1}{4}T_c(M_sl)^2\tau\cot\left[\frac{\pi}{2(1 - \tau)}\right]
+ \frac{\rho_s (Q l)^2}{2} \quad ,
\\ [3mm]
&& F_2 = \frac{ T (M_s l)}{2\pi}
\int_{-B}^{B} \rd \theta \cosh\theta \left[\epsilon(\theta) l \right] \quad ,
\eea
$F_2$ being related to the ground state energy of the solitonic Fermi sea.
(The solitons have a relativistic dispersion relation and usually their 
spectrum is parametrized in terms of the rapidity $\theta$, in the form
$p_0=M_s \sinh \theta$, $\epsilon_0(\theta)=M_s \cosh\theta$, 
for the bare values of energy and momenta. $B$ and  the renormalized value of 
$\epsilon(\theta)$ can be found in \cite{zam}).
Therefore $F_2 \neq 0$ only at $Q > Q_c$. 
In the commensurate phase there will be also a contribution from the the first
part of $F_1$ due to the $Q$-dependence of  $M_s$:
\bea
&&\frac{\phi_0}{2\pi ld}M_{||} = -\frac{\p F}{\p Q} = 
- \left[\rho_s Q_c(0)\right] \frac{Q}{Q_c(0)}+
\nonumber\\[3mm]
&&\frac{T_c}{4}\left\{ n_{\rm sol} + (Ql/2)\frac{\tau}{1 -\tau}
\cot\left[\frac{\pi}{2(1 - \tau)}\right](M_sl)^2\right\} \quad , \\[3mm]
&& \hspace{15mm} n_{\rm sol} = - \frac{\p{\cal E}_0}{\p h} \quad , 
\eea
with $n_{\rm sol}$ being the soliton density. In the limiting case $T \rightarrow 0$
the subcritical contribution to the magnetization is given by 
\bea
\frac{\phi_0}{2\pi dl} M_{||}(Q < Q_c; T=0) = \frac{\pi T_c}{256}\frac{Q}{Q_c}(lQ_c)^3 \quad , 
\eea
where $Q_c$ is given by Eq.(\ref{t0}).

\begin{figure}
\unitlength=1mm
\begin{picture}(80,55)
\put(-1.5,3){\epsfig{file=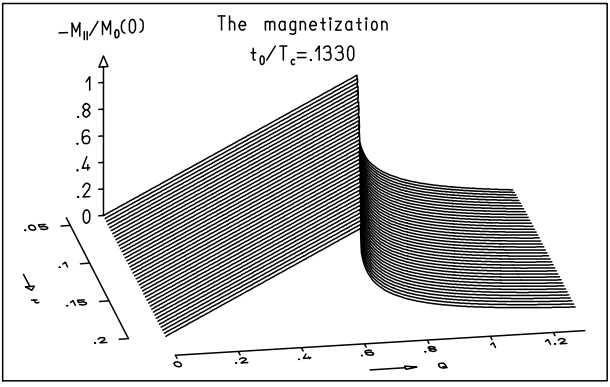,height=50mm}}
\end{picture}
\caption{
Plot of the magnetization on both the C and I phases of the double-layer QH system as a function 
of the magnetic field $Q$ and temperature $(\tau=T/T_c)$ is shown. In the C phase the magnetization 
increases with Q until it reaches the C-I transition critical point. The presence of the soliton 
condensate in the I phase results in a fast decrease of the magnetization.
}
\label{}
\end{figure}

On Fig.~2 we give the dependence of the magnetization on the magnetic field and temperature.
As can be also observed in the figure the presence of the solitons leads
to a decrease of the magnetization of the system. It takes biggest value at small 
temperatures and close to the critical line. For high fields the magnetization
 vanishes.
Close to the critical line, in the I phase the thermal fluctuations  of the solitons 
renormalize the magnetization decrease by changing its behavior from  $1/\ln(1/\epsilon)$ 
as $T \rightarrow 0$ to $\sqrt{\epsilon}$ at finite temperatures, where $\epsilon=[\tau/\tau_c(Q)-1$].

In the following we give also the asymptotic value of the density of solitons (or the soliton 
lattice wave number $Q_s=2\pi/L_s=2\pi n_{\rm sol}$, $L_s$ is the distance of solitons from each other).
This is related to the soliton lattice magnetization by $M_{||}= (4\pi^2 \rho_s/\phi_0) n_{\rm sol}$,
where $n_{\rm sol}$ is the density of the solitons in the system in the IC phase and $\phi_0$
is the magnetic flux quantum.
The density of solitons can be found as:
\bea
n_{\rm sol} = \frac{1}{L_s}=-\frac{\p F_2}{\p {\bar h}}= - \frac{{\bar \beta}^2}{2 \pi}
\frac{\p F_2}{\p Q}=-\frac{\p  {\cal E}_0}{\p h} \quad .
\label{n_sol}
\eea
${\bar h}=2\pi Q/{\bar \beta^2}$ is the field coupled to the
soliton charge, or the number of particles, and
${\bar \beta}^2=1/\rho_s$, [see Eq.~(\ref{sg-class})].
At finite temperature we find
\bea
&& [Q  \rightarrow  Q_c\, , \, T\neq 0]
\nonumber \\[3mm]
&& \frac{Q_s}{Q}= \frac{8\sqrt{2}\, \tau\,  M_s^2}{Q Q_c}\, \epsilon^{1/2}\,
\left[1-\frac{16}{3\sqrt{2}}G(0) \, \epsilon^{1/2}+\ldots \right] -
\nonumber \\[3mm]
&&\hspace{30mm}
- \frac{8\sqrt{2}}{3} \, \frac{\tau}{1-\tau} \, l^2 M_s^2  \, \epsilon^{3/2}+\ldots \quad .
\eea
The last term is a result of the fact that the tunneling amplitude is field dependent.

Similarly along the fixed $Q$ line, $Q_s$, to first order in $(\epsilon_\tau)^{1/2}$, has square root
behavior in $T$
\bea
Q_s = \frac{8\sqrt{2}\, \tau\,  M_s^2}{Q_c}
\sqrt{- \tau_c \frac{Q'_c}{Q}}\left(\frac{\tau}{\tau_c}-1\right)^{1/2}
+ \ldots \quad .
\eea
For smaller temperatures (fixed $\tau$) the square root dependence holds only in a
decreasingly small
interval $Q$ above $Q_c$. At these temperatures
for larger $Q$ the dependence of $1/L_s$ on $Q$ changes
Using the expression for the soliton free energy \cite{papa} we find
\bea
\nonumber
&& [Q\rightarrow Q_c \; ,\; (T\rightarrow 0)]
\\[3mm]
&& \frac{Q_s}{Q}=-\frac{8 \pi^2 \tau^2 \tilde{M_s}^2
e^{-(Q l/2)^2} }{Q Q_c} \left[ \frac{(\ln\epsilon-1)}
{\ln^2\epsilon} + \right.
\\[3mm]\nonumber
& & \hspace{20mm}
\left.
 + \frac{Q_c Q l^2}{2} \, \frac{ \epsilon}{\ln(1/\epsilon)} \right]
\quad ,
\eea
where $\tilde{M_s}$ would be the soliton mass in the case the tunneling amplitude
is independent of the magnetic field $Q$.

 On Fig.~3 we give the plot of the specific heat for $\mu = 0.03$. 
At small $T$ in the C phase the specific heat per unit area is linear in 
temperature:
\bea
C_{\rm V}/T = l^{-2}R \frac{\pi}{8}\mu \ln^2\mu \quad . 
\eea
For $\mu =0.03$ we have $\gamma l^2 \approx 0.14R$. 

\begin{figure}
\unitlength=1mm
\begin{picture}(75,52)
\put(1,3){\epsfig{file=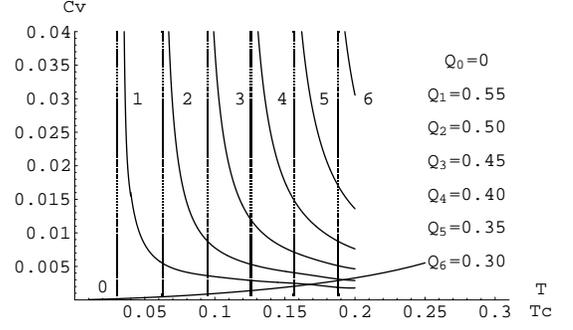,height=45mm}}
\put(15.,8.){\dashbox{0.2}(0,35)}
\put(21.,8.){\dashbox{0.2}(0,35)}
\put(27.,8.){\dashbox{0.2}(0,35)}
\put(32.8,8.){\dashbox{0.2}(0,35)}
\put(38.5,8.){\dashbox{0.2}(0,35)}
\put(44.4,8.0){\dashbox{0.2}(0,35)}
\end{picture}
\caption{
The total specific heat of the
double layer QH system for different values of the magnetic field $Q$.
The presence of the soliton condensate
leads to a decrease in the specific heat, except close to the critical line where it
diverges. As $Q$ increases the critical point goes towards $T=0$ point and the divergence changes form
from $1/\sqrt{\epsilon}$ to $1/[{\epsilon \ln(1/\epsilon})]$. 
Here the tunneling amplitude is $t=0.1$ meV.} 
\label{CV}
\end{figure}

In the I phase the soliton condensate leads to a decrease of the specific heat,
 except close to the critical line where it diverges. It is worth mentioning that
the leading order term on the condensate contribution to the specific heat
changes form from $1/[\epsilon \ln(1/\epsilon)]$ when $T \rightarrow 0$ to $1/\sqrt{\epsilon}$ at 
finite temperatures. 

In the following we calculate the compressional stiffness element $K_{11}$ of the stiffness
tensor $K_{ij}$. The plot which is given below applies in the case of the bilayers up to temperature
$T\approx 3 T_{\rm KT}$, and for other systems analogues to SG, in the whole temperature region.
We calculate $K_{\rm 11}$ as the change of the free energy by varying the
spacing between solitons.
This is achieved by calculating the second derivative with respect to the component
${\bf Q}_{s1}$ for fixed ${\bf Q}$
\bea
K_{11}(Q,\tau) = \frac{\p^2 f_{\rm sol}}{\p Q_{\rm s1}^2}
\quad ,
\eea
where $f_{\rm sol}$ is the soliton contribution to the free energy and is equal to $F_2$.
Substituting
\bea
Q_{s1} = -\frac{1}{\rho_s}\frac{\p f_{\rm sol}}{\p Q}
\quad ,
\eea
one finds for $K_{11}$ the following expression
\bea
K_{11} & = & -\rho_s^2 \left[\frac{1}{{f_{\rm sol}}_{QQ}}-\frac{{f_{\rm
sol}}_Q {f_{\rm sol}}_{QQQ}}{{f^3_{\rm sol}}_{QQ}}\right]
\nonumber \\[3mm]
&=& \rho_s \frac{1}{\left(\p Q_s / \p Q \right)}\left[1 -
\frac{Q_s \, (\p^2 Q_s / \p Q^2)}{(\p Q_s / \p Q)^2}\right]
\quad .
\eea
Note that in Ref.~\cite{hanna} the second term is not taken into account.

\begin{figure}
\unitlength=1mm
\begin{picture}(80,55)
\put(-1.5,3){\epsfig{file=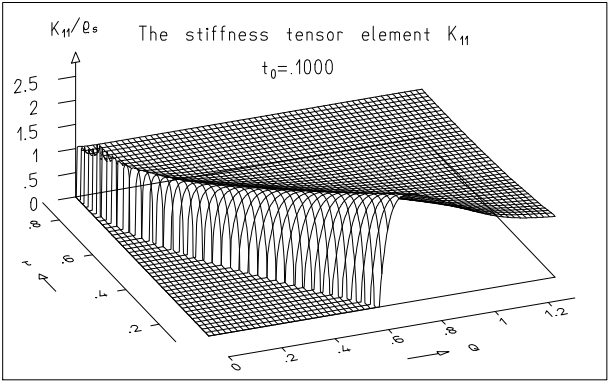,height=50mm}}
\end{picture}
\caption{
Plot of the temperature and field dependence of the compressional
stiffness tensor element $K_{11}$ is shown. Close to $T_c$,
$K_{11}$ approaches  the constant value
$\rho_s$. As $T\rightarrow T_{\rm KT}$, $K_{11}$ reaches quickly values close to the asymptotic
value, $\rho_s$, and does not vary much with the magnetic field.}
\label{stiffness}
\end{figure}

We calculate $K_{11}$ numerically and give in Fig.~\ref{stiffness}
its dependence as a function of temperature and magnetic field.

In the following we give the asymptotic behavior of $K_{11}$, keeping
only the leading order term.

At finite temperatures, near the critical
line, $K_{11}$ increases as a square root function of the magnetic field

\bea
& & [Q  \rightarrow  Q_c\, , \, T\neq 0]
\nonumber \\[3mm]
& & K_{11}=\rho_s \frac{Q_c^2}{4\sqrt{2} \tau M_s^2} \epsilon^{1/2}
+\ldots \quad .
\eea

As it can be seen on Fig.~\ref{stiffness},
the compressional elastic constant characterizing the soliton lattice,
reaches quicker its asymptotic value  at higher temperatures than
at lower ones.
At temperatures close to $T_c$, $K_{11}$  has values
very close to $\rho_s$ and changes very little for variations of the magnetic
field.

For temperatures towards zero the square root holds in a decreasingly small
interval $Q$ above $Q_c$. For larger $Q$, outside this interval,
the dependence of $K_{11}$ on $Q$ changes.
At zero temperatures the $Q$ interval of the square root behavior
vanishes and the $Q$ dependence of $K_{11}$ is
\bea
& & [Q  \rightarrow  Q_c\, , \, T\rightarrow 0\,]
\nonumber \\[3mm]
& & K_{11}=\rho_s \frac{Q_c^2}{8\pi^2\tau^2 M_s^2}\, \epsilon \,
\ln^3(1/\epsilon) + \ldots \quad .
\eea

The large $Q$ asymptotics at zero temperature is given by
\bea
& & [Q  \rightarrow  +\infty \, , \, T\rightarrow 0\,]
\nonumber \\[3mm]
& &
\label{stiff}
K_{11} = \rho_s\left[ 1 + \frac{}{}
\right.
\label{QinftyT0}
\\[3mm] \nonumber
& & \left. +
\frac{1}{64} \left(    \frac{\pi}{2} \frac{\tilde{Q}_c}{Q}    \right)^4
\frac{ (18 + 9 Q^2 l^2 +2 Q^4 l^4 +Q^6 l^6) }
{e^{(Q l)^2/2}}+\ldots \right] \quad .
\eea
The asymptotics for the case of the tunneling amplitude independent of
the magnetic field is obtained by substituting $l=0$.

The large $Q$ asymptotics for finite temperatures and field independent
tunneling is given by
\bea
& & [Q  \rightarrow  +\infty \, , \, T\neq 0\,]
\\[3mm] \nonumber
& &
K_{11} = \rho_s\left[1 +
\tilde{\alpha}(\tau) \frac{\left(18 -84 \tau + 128 \tau^2_{} -64 \tau^3\right) }{
Q^{4 - 4\tau} } + \ldots \right] \, ,
\eea
where
\bea
\tilde{\alpha}(\tau) = &&
 \frac{\Gamma(\tau)}{\Gamma(-\tau)}
\frac{\Gamma(5/2-\tau)}{\Gamma(1/2+\tau)}
\frac{({\tilde{m}^*})^{4(1-\tau)}}{(3-2\tau) (2\tau - 1)}
\quad .
\eea
It is not the soliton mass itself, but the combination \cite{papa2}
\bea
\nonumber
m^* = 2\sqrt\pi \frac{\Gamma\left[\frac{1}{2(1 - \tau)}\right]}
{\Gamma\left[\frac{\tau}{2(1 - \tau)}\right]}lM_s = 
4\left[\mu_0 e^{-(Ql)^2/4}\frac{\Gamma(1 - \tau)}{\Gamma(1 + \tau)}\right]^{
\frac{1}{2(
1 - \tau)}}
\eea
which enters as a scale in the expansion above.

The case with the tunneling dependent on the magnetic field
$K_{11}$ has the following asymptotic behavior
\bea
& & [Q  \rightarrow  +\infty \, , \, T\neq 0\,]
\nonumber \\[3mm]
& &
K_{11} = \rho_s\left[1 + \alpha(\tau) \frac{l^6 Q^{2+4\tau}}{e^{(l Q)^2 /2} }
+\ldots \right] \quad .
\label{Qinftylneq0}
\eea
This is consistent with (\ref{QinftyT0}), if in (\ref{Qinftylneq0}) $\tau=0$ is
substituted. In (\ref{Qinftylneq0}) only the highest order term is kept.

In this work we made use of the exact solution of the sine-Gordon model to extend the previous
results for the bilayer QH system \cite{hanna} to finite temperatures by including the thermal 
fluctuations of the soliton lattice.
We calculated the magnetization and its temperauture dependence in both C and I phases. 
In addition we calculated the specific heat of the bilayer system which could not be achieved with the 
method of the previous work \cite{hanna}. 
The presence of the condensate results in a decrease of both of  the 
specific heat of the system (except close to the critical line where it diverges) and its magnetization. 

We observe however that this method cannot be applied in the whole temperature interval where the results
for the sine-Gordon model are available.
Above the Berezinskii-Kosterlitz-Thouless transition temperature, in the interval
$(T_{\rm BKT},T_c)$ where $\cos\beta\varphi$ and $\cos(2\pi \vartheta/\beta)$ are both relevant, 
the system is in a new phase with coexisting order parameters. 
(As we argued we could extend the interval of SG applicability, with decreasing accuracy,
 up to $3T_{\rm BKT}$).
In this phase  this system can no longer be described by Eq.~(\ref{sg-class}). 
Furthermore, as pointed out by \cite{macd2}, at higher temperatures the role of the spin degree of freedom 
becomes important and the model used would need to be modified.

We notice finally that different authors have found, instead of $T_C$, the Berezinskii-Kosterlitz-Thoouless 
transition temperature as the point where the soliton lattice melts down.

The authors are grateful to A.H. MacDonald for interesting discussions.
 A.M.T.  is supported by the US DOE under
contract number DE-AC02-98 CH 10886 and EPSRC grant number 99307266.
E.P. is supported in part by Welch Foundation and 
by the National Research Foundation under grant DMR 0115947.

\end{document}